# Fluorescent redox-dependent labeling of lipid droplets in cultured cells by reduced phenazine methosulfate

Juan C. Stockert[1,2,*], María C. Carou[1], Adriana G. Casas[3], María C. García Vior[4], Sergio D. Ezquerra Riega[4], María M. Blanco[4], Jesús Espada[5,6], Alfonso Blázquez-Castro[7], Richard W. Horobin[8], Daniel M. Lombardo[1,*]

[1] *Universidad de Buenos Aires, Facultad de Ciencias Veterinarias, Instituto de Investigación y Tecnología en Reproducción Animal, Cátedra de Histología y Embriología, Buenos Aires C1427CWO, Argentina*

[2] *Universidad de Buenos Aires, Instituto de Oncología "Angel H. Roffo", Area Investigación, Buenos Aires C1417DTB, Argentina*

[3] *Centro de Investigaciones sobre Porfirinas y Porfirias, Hospital de Clínicas, Universidad de Buenos Aires, and CONICET, C1120AAF, Argentina*

[4] *Universidad de Buenos Aires, Facultad de Farmacia y Bioquímica, Departamento de Química Orgánica, C1113AAD, CABA, Argentina*

[5] *Experimental Dermatology and Skin Biology Group, Ramón y Cajal Institute for Health Research, Ramón y Cajal University Hospital, 28034 Madrid, Spain*

[6] *Centro Integrativo de Biología y Química Aplicada (CIBQA), Universidad Bernardo O´Higgins, Santiago 8370854, Chile*

[7] *Departamento de Biología, Facultad de Ciencias, Universidad Autónoma de Madrid, Madrid 28049, Spain*

[8] *Chemical Biology and Precision Synthesis, School of Chemistry, University of Glasgow, Glasgow G12 8QQ, Scotland, UK*

\* Corresponding authors: E-mail addresses: jcstockert@fvet.uba.ar (J.C. Stockert), dlombard@fvet.uba.ar (D.M. Lombardo).

**ABSTRACT**

Natural and synthetic phenazines are widely used in biomedical sciences. In dehydrogenase histochemistry, phenazine methosulfate (PMS) is applied as a redox reagent for coupling reduced coenzymes to the reduction of tetrazolium salts into colored formazans. PMS is also currently used for cytotoxicity and viability assays of cell cultures using sulfonated tetrazoliums. Under UV (340 nm) excitation, aqueous solutions of the cationic PMS show green fluorescence ($\lambda$em: 526 nm), whereas the reduced hydrophobic derivative (methyl-phenazine, MPH) shows blue fluorescence ($\lambda$em: 465 nm). Under UV (365 nm) excitation, cultured cells (LM2, IGROV-1, BGC-1, and 3T3-L1 adipocytes) treated with PMS (5 µg/mL, 30 min) showed cytoplasmic granules with bright blue fluorescence, which correspond to lipid droplets labeled by the lipophilic methyl-phenazine. After formaldehyde fixation blue-fluorescing droplets could be stained with oil red O. Interestingly, PMS-treated 3T3-L1 adipocytes observed under UV excitation 24 h after labeling showed large lipid droplets with a weak green emission within a diffuse pale blue-fluorescing cytoplasm, whereas a strong green emission was observed in small lipid droplets. This fluorescence change from blue to green indicates that reoxidation of methyl-phenazine to PMS can occur. Regarding cell uptake and labeling mechanisms, QSAR models predict that the hydrophilic PMS is not significantly membrane-permeant, so most PMS reduction is expected to be extracellular and associated with a plasma membrane NAD(P)H reductase. Once formed, the lipophilic and blue-fluorescing methyl-phenazine enters live cells and mainly accumulates in lipid droplets. Overall, the results reported here indicate that PMS is an excellent fluorescent probe to investigate labeling and redox dynamics of lipid droplets in cultured cells.

*Keywords*

Fluorescent labeling

Lipid droplets

PMS

QSAR

Redox reagents

**Short title:** Fluorescent lipid staining by reduced PMS

**Introduction**

Phenazine methosulfate (PMS) was first synthesized by Kehrmann and Havas (1913) [1] from phenazine and dimethylsulfate. Later it was applied in dehydrogenase histochemistry [2, 3, 4, 5], and studies on the photosynthesis of bacteria [6].

The hydrophilic cation PMS is now widely used as an intermediate redox reagent for coupling reduced dehydrogenase coenzymes to the reduction of tetrazolium salts into colored formazans [7, 8, 9, 10, 11, 12, 13, 14]. A similar electron-transfer agent, 1-methoxy-PMS, has also been used for revealing dehydrogenase activity in living hepatocytes [15]. Other similar but seldom used redox intermediates are Meldola blue, menadione, and pyocyanin.

In addition, PMS is also applied in cytotoxicity and viability assays of cell cultures, using sulfonated tetrazoliums such as MTS, WST-1, WST-8, and XTT [16, 17, 18, 19, 20]. PMS also accelerates the reduction of MTT and CTC tetrazolium salts in viability assays [21]. Toxic effects of PMS on living cells cannot be ruled out since this compound inhibits oxidative phosphorylation [22], and is mutagenic in *Escherichia coli* and *Salmonella typhimurium* assays [23].

The yellow PMS is capable of oxidizing NADH, passing electrons to tetrazolium salts [5]. PMS is reduced by flavoproteins (succinic dehydrogenase), and non-enzymatically by NAD(P)H, ascorbic acid, vitamin K, reduced ubiquinones, dithionite, and sodium borohydride [24]. The reduced PMS (methyl-phenazine, MPH) is colorless and easily reoxidized by oxygen or via the respiratory chain, and the reaction is blocked by inhibitors of cytochrome oxidase such as cyanide ions [22]. Methyl-phenazine is hydrophobic and has low solubility in aqueous media [12]. The chemical structures of PMS and methyl-phenazine are shown in Fig. 1.

Interestingly, both natural and synthetic phenazines have important applications in biomedical sciences [25]. Examples include the blue pigment pyocyanin from *Pseudomonas aeruginosa* [26], and several azine dyes such as Janus green B, neutral red, and safranine O [27]. There are also phenazine derivatives with antibacterial, antileukemic, antimycotic, antiparasitic, antitubercular, and antitumoral activities, as well compounds with immune-suppressive, and biofilm-eradicating properties [28, 29, 30, 31, 32, 33, 34]. Massive accumulation of crystal-like inclusions of the antileprosy phenazine, clofazimine, within live macrophages have been described [35]. In a similar way to what occurs with some oxidized and reduced thiazine dyes [36], the redox PMS-methyl-phenazine cycling activity could be partially responsible of cytotoxic and antitumoral effects of some phenazine derivatives.

According to traditional color theory, PMS is not a dye as it lacks an auxochrome group [37], and perhaps as a consequence, its fluorescence properties and possible interactions with biological substrates within living cells have been largely overlooked. A few exceptions to this can be cited. A chemiluminescence flash induced by PMS in the presence of $H_2O_2$ and reductants (NADH, ascorbic acid) was observed instrumentally, possibly involving the green semiquinone PMS cation [38]. Spectroscopic studies on PMS absorption are known [26, 39, 40], but emission properties and possible uses of PMS and derivatives as labeling probes have been overlooked in fluorescence microscopy.

Here we report a previously unnoticed, unexpected redox-dependent bright blue fluorescence of cytoplasmic granules, corresponding to lipid droplets, in different cultured cell lines after treatment with PMS. This constitutes an experimental basis for using this compound as a novel fluorescent probe to study the labeling and redox dynamics of lipid droplets.

**Material and Methods**

1. Chemicals

PMS (5-methylphenazinium methyl sulfate, phenazine methosulfate) was purchased from Sigma-Aldrich (P9625, St. Louis, USA; purity: 98%), and used as received. Milli-Q water was obtained from a Milli-Q system (Millipore). Zinc powder, acetic acid, and n-octanol were obtained from Sigma–Aldrich (Darmstadt, Germany). The copolymer Tetronic® 1107 (Ethylenediaminetetrakis (propoxylate-block-ethoxylate)-tetrol-poloxamine) was obtained from BASF (MW: 15 kDa, New Jersey, USA). The reduction of aqueous PMS solutions to methyl-phenazine was accomplished using Zn powder at acid pH. PMS (2 mg, 0.0065 mmol) was dissolved in Milli-Q water (4 mL) previously acidified with acetic acid (0.1 mL). Finely powdered Zn° (10 mg, 0.153 mmol) was added slowly at room temperature (RT). After complete addition, the mixture was stirred at RT until the yellow color disappeared (0.5 h), then filtered under argon, extracted with n-octanol (3 x 3 mL), and the organic phase immediately used to prepare a dilution in order to record the emission spectra. The solution obtained was maintained in the argon atmosphere until all the measurements were made. The polymeric micelle solution (10% w/v) was prepared by dissolving the required amount of copolymer T1107 in milli-Q water (pH 7–8) at 4°C followed by the equilibration of the system at 25° C for at least 24 h before use [41], in order to obtain the emission spectra of methyl-phenazine in the lipophilic T1107 micelles.

2. Spectroscopy

Emission spectra were obtained using fluorescence spectrofluorometers LS55 (Perkin Elmer, Waltham, USA), and PTI Quanta Master QM4 ((PhotoMed GmbH, Seefeld, Germany), using a 10x10 mm quartz cuvette. All experiments were performed at RT. The emission spectra of aqueous PMS and methyl-phenazine in water, n-octanol and T1107 micelles were collected at an excitation wavelength of 340 nm and recorded between 350 nm and 750 nm.

To estimate lipophilic/hydrophilic properties, the logarithm of the octanol-water partition coefficient (log *P*) of PMS and methyl-phenazine was determined by the traditional shake-flask method at RT [42]. MilliQ water and n-octanol were mutually saturated. After saturation, a known amount of drug was loaded into the aqueous phase such that the concentration of the final dilutions lies in the range of developed analytical UV method. Equal amounts of a drug-loaded aqueous phase and n-octanol were shaken together on a mechanical shaker for 30 minutes, centrifuged at 3000 rpm for 15 min to afford complete phase separation, and the n-octanol phase was removed. The absorbance of the water phase was measured spectrophotometrically at the corresponding $\lambda_{max}$. Also, log $P$ values were calculated by using two different fragment methods, namely those of Hansch and Leo [43] and Black and Mould [44]. Log $P$ values were then inserted into predictive QSAR models [45, 46, 47] to evaluate probable cellular uptake and intracellular localization of both PMS and methyl-phenazine. Note that this procedure used the log $P$ values provided by the Hansch and Leo method [43], as the QSAR models required log $P$ values only available from this system.

3. Cell cultures and treatments

The cultured cells, LM2 (murine mammary adenocarcinoma, Instituto Roffo, Buenos Aires, Argentina), IGROV-1 (human ovary carcinoma, ATCC), BGC-1 (bovine ovarian granulosa), and activated mouse 3T3-L1 adipocytes were used for microscopic labeling assessment by PMS. In the case of BGC-1, both cells and follicular fluid were obtained by puncturing ovarian follicles 3-8 mm in the size of cows and heifers after slaughter. Cells from the primary BGC-1 culture were grown on 18×18 mm coverslips in multiwell plates in culture medium DMEM+F12 (pH 7.4) supplemented with 5% fetal calf serum, 2 nM L-glutamine, and antibiotics as previously described [48]. Cells from passage 8 were used for PMS labeling. Murine preadipocyte 3T3-L1 fibroblasts were differentiated to adipocytes by induction for 5 days after seeding with 0.1 µM dexamethasone, 0.5 mM isobutyl methyl xanthine (IBMX; Sigma I-7018), and 2 µM insulin (Bovine; Sigma I-5500), followed by additional treatment with insulin alone for 3 days. The induction procedure was performed following protocols slightly modified [49, 50].

All cell cultures were treated with PMS (from 5 to 20 µg/mL for 15 min to 3 h), washed, and mounted with PBS. Observations were made using fluorescence microscopes Olympus BX51 (Olympus, Tokyo, Japan), and Leica DM4000B Led (Wetzlar, Germany), under ultraviolet (UV, 365 nm) and blue (436 nm) exciting light. To test the possible reduction to methyl-phenazine directly induced by the culture medium, PMS at a final concentration of 50 µg/mL was added to the complete culture medium (at 37 ºC, in the absence of cells and pH indicator) and observed 30 min later. In some cases, cell cultures were fixed with 4% aqueous formaldehyde before PMS treatment. To confirm the lipid nature of labeled droplets, formaldehyde-fixed 3T3-L1 adipocytes were stained with 0.36% w/v oil red O (Sigma-Aldrich) solution in 40% v/v isopropanol for 60 min after PMS treatment. The samples were washed in distilled water, counterstained with Mayer's hematoxylin for 10 min, washed with tap water, and mounted in a drop of gum Arabic solution (Winsor & Newton, London, UK). In other cases, adipocytes were allowed to grow for an additional 24 h in complete medium and then observed under UV and blue excitation.

Colorimetric viability assays [51, 52] were performed by triplicate on nearly confluent LM2 cells (24 h post PMS incubation at different concentrations for 30 min), using 50 µg/mL MTT for 3 h, followed by formazan extraction with DMSO, and reading at 560 nm in an Epoch microplate spectrophotometer (Biotek, Winooski, USA). Cell survival was expressed as the percentage of formazan absorbance of treated cells in comparison with that of control cells.

**Results**

The emission spectra of PMS and methyl-phenazine are shown in Fig. 2. Under 340-nm excitation, PMS in distilled water showed a green fluorescence with an emission peak at 526 nm, whereas the fluorescence of methyl-phenazine was blue with emission peaks at 465, 466, and 469 nm in water, T1107 and n-octanol, respectively. After reducing PMS to methyl-phenazine, a large blue-shifted absorption (61 nm) band is observed as consequence of the conjugation disruption present in the molecule (not shown). It must be noted that non-normalized emission spectra of samples with similar concentration (around 3 μM) show an increment of almost 10 times in the fluorescence intensity of methyl-phenazine dissolved in a lipophilic solvent as compared with PMS in water (not shown).

The hydrophilic/lipophilic comparison of PMS and methyl-phenazine is indicated by the experimental and calculated log $P$ values shown in Table 1. Although the log $P$ values for each compound vary with the method of estimation, nevertheless, the oxidized PMS cation is consistently more hydrophilic than the reduced derivative, methyl-phenazine.

Table 1. Log $P$ values of PMS and methyl-phenazine obtained using different methods. Note the consistently negative (hydrophilic) and positive (lipophilic) values of PMS and methyl-phenazine, respectively.

| Log $P$ values | PMS | Methyl-phenazine |
|---|---|---|
| Experimental | -1.22 | 0.16 |
| Calculated by Hansch and Leo (1995) procedure [43] | -1.4 | 3.3 |
| Calculated by Black and Mould (1991) procedure [44] | -3.73 | 2.46 |

The QSAR models predict [45, 46] that the hydrophilic PMS cation is not membrane-permeant, and therefore it is expected to remain outside the plasma membrane. However, the uncharged and lipophilic methyl-phenazine is predicted to enter into live cells and localize in lipid components, in particular lipid droplets and the Golgi apparatus. Since methyl-phenazine is non-ionic, has a log $P < 5$, and is not amphiphilic, significant accumulation in other cell membranes is not expected.

Viability assays were carried out on LM2 cells using the MTT method. With a 30 min incubation, a dose-dependent loss of viability occurred at 24 h when the PMS concentration exceeded 5 µg/mL (Fig. 3).

Under UV exciting light, fluorescence microscopy of all cell cultures subjected to PMS treatments showed a bright blue granular labeling pattern within the cytoplasm (Fig. 4). Generally, some weak and diffuse blue fluorescence was also found in the cytoplasm, but see below; nuclei were non-fluorescent. When observed under bright-field illumination, lipid droplets appeared as highly refrigent vacuoles within the cytoplasm of both control and PMS-treated adipocytes (Fig. 5, A). As shown by fluorescence microscopy after PMS treatment, lipid droplets were the only cell component with a strong and selective blue labeling (Fig. 5, B, C). In 3T3-L1 cells stained with oil red O, this component showed a red color, which indicates its lipid character (Fig. 5, D). Colocalization studies of the methyl-phenazine signal and fluorescent lipid probes such as Nile red and oil red O were not made in living cells because excitation of these probes by the blue emission of methyl-phenazine could generate misleading results. The Golgi apparatus of LM2, IGROV-1, and BGC-1 cells was also slightly labeled, but in 3T3-L1 adipocytes, the intense labeling of large lipid droplets hindered the precise observation of this organelle.

UV-induced blue fluorescence of lipid droplets was clearly observed with all PMS concentrations and incubation times used, although 5 µg/mL and 30 min were the optimal conditions for cell viability and bright selective labeling of lipid droplets. Some fading was observed after PMS labeling, which did not impede the detailed observation of cells. No blue fluorescence was found in any structure in cells previously fixed with formaldehyde. In addition, no discoloration of the yellow PMS was observed in the presence of culture medium alone *in vitro*. With blue excitation (436 nm), the emission of lipid droplets was pale green. 3T3-L1 adipocytes treated with 5 µg/mL PMS for 30 min and observed 24 h later under UV excitation showed a striking change of emission color in lipid droplets (Fig. 6). In this case, weak green fluorescence was found in large lipid droplets, which were surrounded by a clear blue rim within a diffuse pale blue-fluorescing cytoplasm. Small lipid droplets with a strong green fluorescence were commonly observed in PMS-treated cells (Fig. 6).

**Discussion**

When reduced, several dyes - including methylene blue, Janus green B, and ethidium bromide - become colorless or show changes in their fluorescence emission [27, 53]. Another example is the reduction of the blue and non-fluorescent indigo carmine to the green-fluorescing leuco-compound [54]. The present investigation shows that after reduction, the green emission color of PMS changes to a blue emission from methyl-phenazine, which is the only observed short-term microscopic fluorescence in live cells after PMS treatments.

Regarding possible mechanisms of labeling, as the hydrophilic PMS is expected to be plasmatic membrane non-permeant, most PMS will remain in the cell culture medium. In this case, as extracellular PMS reduction by the culture medium itself does not occur, reduction is likely to be associated with biochemical processes at the cell surface of living cells, plausibly due to a plasma membrane-bound NAD(P)H reductase [55, 56], which has been proposed to be the NAD(P)H: quinone oxidoreductase 1 (NQO1) [36]. The resulting lipophilic and blue-fluorescing methyl-phenazine will then enter into live cells by passive diffusion through the plasma membrane, followed by accumulation within lipid droplets and the Golgi apparatus (Fig. 7). Note that on account of the strong reducing power of the cytoplasm [53, 57], which is due to the abundance of reduced NAD(P)H and dehydrogenases, methyl-phenazine will remain reduced, providing suitable stability for the selective blue fluorescence of lipid structures.

However, 24 h after PMS labeling of 3T3-L1 adipocytes, the emission color of lipid droplets had changed from blue to green, indicating that reoxidation to PMS can occur. It is tempting to speculate that the gradient of blue-to-green color change shown by lipid droplets is related to a decrease of reducing power or to an increase of oxidative stress, perhaps even driven by PMS-methyl-phenazine redox cycling (see Fig. 3 for MTT cell viability assay). Reducing conditions are present in the cytoplasm, while molecular oxygen dissolves better in lipophilic environments, like the core of the lipid droplets. Therefore, a PMS-methyl-phenazine loaded lipid droplet redox cycle can be formed, in which PMS remains reduced to methyl-phenazine at the droplet-cytoplasm interface, and then the reduced product diffuses to the core where it is reoxidized to PMS by electron transfer to molecular oxygen. Tentatively, this can lead to reactive oxygen species (ROS) generation at the lipid droplet core. The ultimate fate of these ROS is speculative, but it could explain the observed decrease in cell viability at high PMS concentrations (see Fig. 3). Ongoing research is devoted to ROS detection after PMS cell incubation.

QSAR modeling of cell uptake and intracellular localization supports these conclusions. According to QSAR models [46] the sulfonated tetrazolium salts MTS, WSR-1, and XTT are hydrophilic and membrane-impermeable anions which do not enter live cells [58]. As reported here, cationic PMS is similarly predicted by QSAR modeling [45] to be membrane non-permeant. Hence, the reduction of PMS must occur outside live cells. In keeping with this, the green fluorescence of PMS is not observed within live cells, except following the intracellular presence of methyl-phenazine. Thus, when PMS is reduced to the blue-fluorescing and lipophilic methyl-phenazine, this compound is predicted to be membrane-permeant and so to passively enter the cell. Further, as its log $P$ value lies between 2-5, methyl-phenazine is predicted to accumulate in lipid droplets and the Golgi apparatus [45, 47]. Accordingly, the most intense staining was that of lipid droplets, with some paler staining of the Golgi apparatus seen. Strong, generalized staining of biomembranes is not predicted, as the lipophilicity of methyl-phenazine is less than 5, and the molecule is not amphiphilic. However, the weak cytoplasmic background fluorescence could be due to small amounts of methyl-phenazine being taken into other membranous organelles and transfer vesicles.

The lipid droplet is a lipid storage compartment and is now considered an important cell organelle. Lipid droplets comprise a hydrophobic core (mainly neutral triglycerides) surrounded by a phospholipid monolayer. Staining of lipid droplets in cultured cells has been recently reviewed [59], and some lipophilic probes such as Nile red, oil red O, chlorophyll-oil micelles, and BODIPY derivatives are commonly used to reveal lipid droplets in fluorescence microscopy [57]. In the case of live cells subjected to MTT viability assays, lipid droplets are also selectively labeled by the lipophilic MTT formazan [52, 58].

Taking into account the current interest in the study of selective fluorochromes for lipid detection [57, 59, 60], the present results indicate that, in addition to such currently available lipid probes, methyl-phenazine can be used as a novel lipid droplet fluorochrome for vital labeling and redox assessing of cultured cells. As a whole, the results reported here are relevant in the biomedical field and may contribute to the development of experimental approaches using PMS as a fluorescent probe to analyze the dynamics of lipid droplets in different redox conditions.

**Acknowledgements**

This work was partially supported by grants from UBACyT 2017-2020, 20020160100062BA, 20020170200301BA, and ANPCyT PICT 1014-0727 (Argentina). The authors thank the valuable assistance of G.M. Di Venosa, D.A. Sáez, and L.M. Luquez. ABC acknowledges funding under the Marie Skłodowska-Curie Action COFUND 2015 (EU project 713366-InterTalentum). RWH thanks Prof Graeme Cooke, University of Glasgow, UK, for the provision of facilities.

**Conflict of interests**

The authors declare that there is no conflict of interests

**Figures**

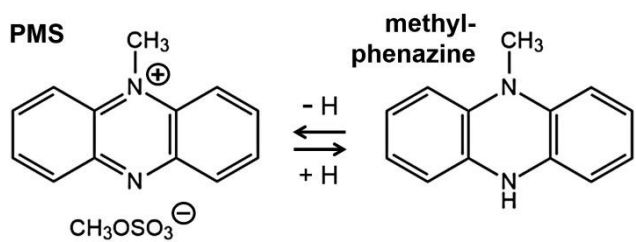

**Fig. 1.** Chemical structures of phenazine methosulfate in its oxidized cationic form (PMS), and the reduced derivative, methyl-phenazine. The incorporation of a hydrogen at the N atom of the central ring in PMS leads to a lower delocalization of the electrons, resulting in a color loss of methyl-phenazine.

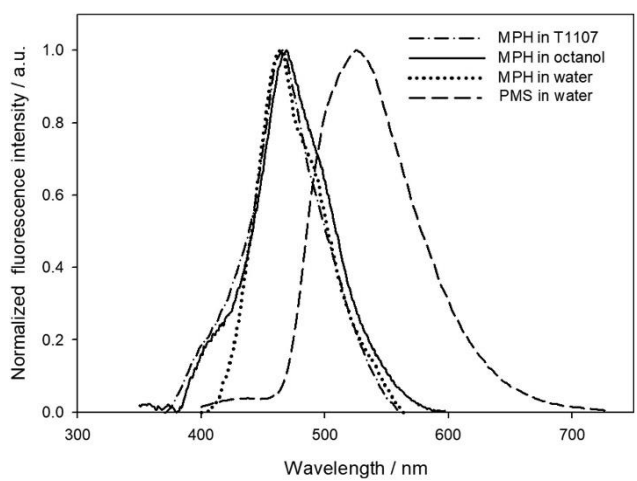

**Fig. 2.** Normalized emission spectra of PMS in water, and methyl-phenazine (MPH) in n-octanol, lipophilic micelles (T1107) and water.

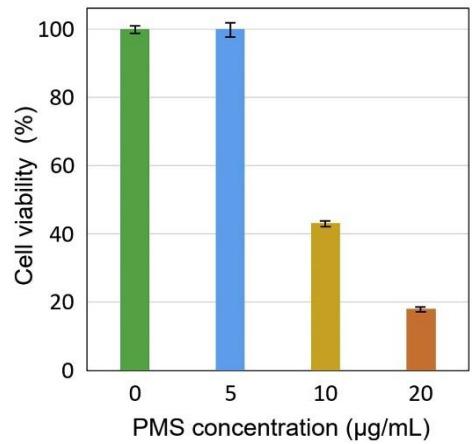

**Fig. 3.** MTT viability at 24 h of LM2 cells following 30 min PMS treatment at different concentrations. Mean ± SD values from at least three different assays are shown.

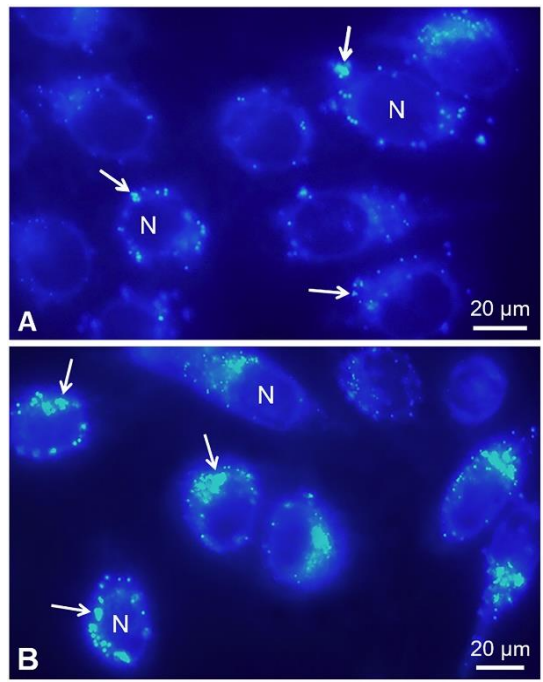

**Fig. 4.** Fluorescence micrographs of cultured LM2 cells (A) and IGROV-1 cells (B) after treatment with PMS (5 µg/mL for 30 min), showing the bright blue emission of lipid droplets (arrows) under 365-nm excitation. N: nuclei.

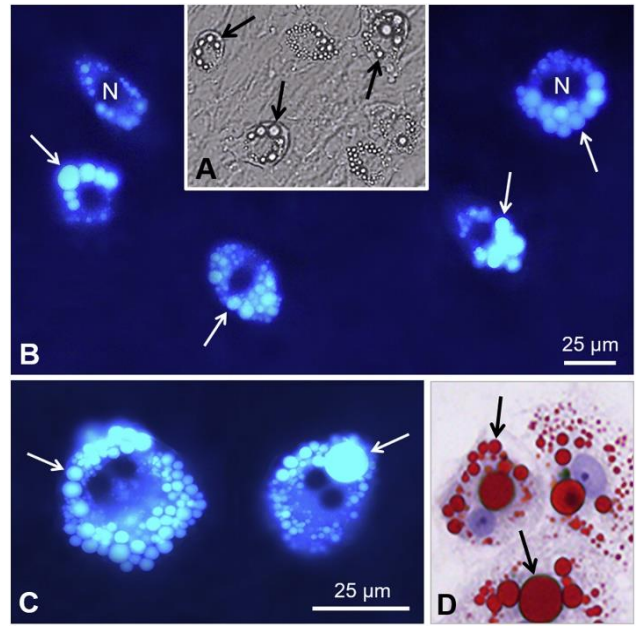

**Fig. 5.** A: unstained 3T3-L1 adipocytes observed under bright-field illumination, showing refringent lipid droplets (arrows). B, C: fluorescence images of 3T3-L1 adipocytes treated with PMS (5 µg/mL, 30 min), showing the selective blue labeling of lipid droplets (arrows). N: nuclei. Note the binucleate cells in C (intracellular dark circles). UV (365 nm) excitation. D: Bright-field image of 3T3-L1 adipocytes fixed in formaldehyde and stained with oil red O, showing the positive staining of lipid droplets (arrows). Mayer' hematoxylin counterstaining.

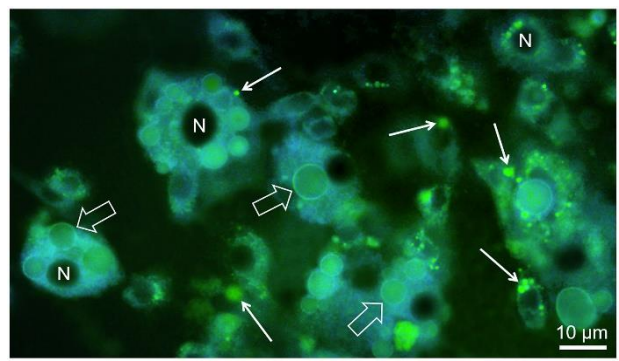

**Fig. 6.** 3T3-L1 adipocytes treated with PMS (5 µg/mL, 30 min), maintained in DMEM for 24 h, and then observed under 365-nm excitation. Within the blue cell cytoplasm,

large lipid droplets show a greenish fluorescent core surrounded by a blue rim (large arrows), whereas small droplets have a bright green emission (small arrows). N: nuclei.

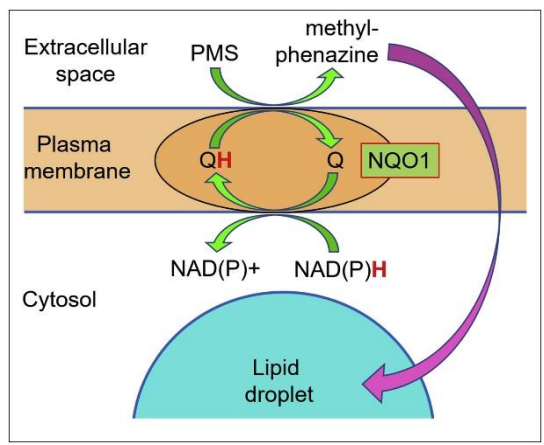

**Fig. 7.** Scheme illustrating the possible molecular reduction mechanism of PMS at the plasma membrane of live cells, and subsequent labeling of a lipid droplet by methyl-phenazine. The reduced coenzyme NAD(P)H provides hydrogen (H) to the coenzyme Q, and then to PMS by the activity of the enzyme NQO1 (NAD(P)H:quinone oxidoreductase 1). In cell viability assays, the extracellular methyl-phenazine reduces sulfonated tetrazolium salts to colored formazans.